\documentclass[lettersize,journal]{IEEEtran}
\IEEEoverridecommandlockouts

\usepackage{cite}
\usepackage{amsmath,amssymb,amsfonts}
\usepackage{algorithm}
\usepackage{algpseudocode}
\usepackage{graphicx}
\usepackage{textcomp}
\usepackage{xcolor}
\usepackage{makecell}
\usepackage{subfig}
\usepackage{multirow}
\usepackage{booktabs}
\usepackage{comment}
\usepackage{graphicx}
\usepackage{setspace}
\usepackage{hyperref}
\usepackage{enumitem}
\usepackage{soul}

\usepackage{tabularx}
\usepackage{array}
\usepackage{booktabs}

\usepackage{booktabs}
\usepackage{pifont}
\usepackage{array}
\usepackage{tabularx}

\usepackage{fontawesome5}
\usepackage{stfloats}
\usepackage{pifont}
\usepackage{xcolor}
\definecolor{UniGreen}{HTML}{228B22} 
\definecolor{UniRed}{HTML}{B22222}   

\usepackage{tikz}
\usetikzlibrary{
    arrows.meta,
    positioning,
    fit,
    shapes,
    shapes.geometric,
    backgrounds
}

\def\BibTeX{{\rm B\kern-.05em{\sc i\kern-.025em b}\kern-.08em
    T\kern-.1667em\lower.7ex\hbox{E}\kern-.125emX}}

\ifodd 1
\newcommand{\com}[1]{\textbf{\color{red} (COMMENT: #1)}} 
\newcommand{\comg}[1]{\textbf{\color{green} (COMMENT: #1)}}
\newcommand{\response}[1]{\textbf{\color{magenta} (RESPONSE: #1)}} 
\else

\newcommand{\com}[1]{}
\newcommand{\comg}[1]{}
\newcommand{\response}[1]{}
\fi
\ifodd 1
\newcommand{\referred}[1]{\textcolor{red}{RefPaper: #1}} 
\else
\newcommand{\referred}[1]{}
\fi

\ifodd 1
\newcommand{\changeblue}[1]{\textcolor{blue}{Modified: #1}} 
\else
\newcommand{\changeblue}[1]{}
\fi

\begin{document}
\bstctlcite{IEEEexample:BSTcontrol}

\title{





SkillComm: Skill-Driven Semantic Communication via Workflow-Aware Token Prioritization






\thanks{The authors are with the Technology and Engineering Center for Space Utilization, Chinese Academy of Sciences, Beijing, 100094, China (e-mails: \{mengziyang24, lulu\}@csu.ac.cn).}



}

\author{
    \IEEEauthorblockN{Ziyang~Meng,~\IEEEmembership{Student~Member, IEEE} and Lu~Lu,~\IEEEmembership{Member, IEEE}} 
}


\maketitle

\begin{abstract}
As wireless visual intelligence evolves from isolated task inference to ordered skill workflows, the core communication bottleneck shifts from transmitting a single semantic representation to coordinating reusable skill states under channel constraints. Existing DeepJSCC and prompt-guided visual transmitters usually treat each task as an independent full-token transmission, with limited reuse of execution memory across semantic workflows. This is inefficient for workflows such as \textit{Detect}$\rightarrow$\textit{Segment}$\rightarrow$\textit{Keypoint}, where later stages mainly require state-relevant semantic updates rather than repeated full-grid transmission.
This letter proposes \textbf{SkillComm}, a skill-driven semantic communication (SemCom) framework that utilizes reusable skill states as communication context to jointly guide workflow-aware token prioritization and memory-assisted token-grid reassembly. {A shared \textit{Skill-Book} first synchronizes the transmitter and receiver by mapping a high-level visual intent into an executable skill sequence. Conditioned on this synchronized workflow,} workflow-aware adaptive token selection exploits cross-step execution memory to prioritize state-active tokens before joint source-channel coding (JSCC) transmission. Finally, the receiver reconstructs a task-ready token grid by merging the decoded active tokens with its local historical memory. 
Evaluations on the MS COCO 2017 validation dataset across the \textit{Detect}$\rightarrow$\textit{Segment}$\rightarrow$\textit{Keypoint} workflow show that workflow-aware token prioritization enables SkillComm to achieve a \textbf{51.2\%} token transmission cost reduction, while retaining \textbf{99.4\%} upper-bound-normalized average precision in the high SNR regime. These results provide quantitative evidence that reusable skill states can drive workflow-aware token prioritization, enabling selective semantic innovation delivery for future agentic and embodied visual intelligence.
\end{abstract}

\section{Introduction}
\label{sec:introduction}

{As wireless visual intelligence evolves from isolated task inference to ordered skill workflows, semantic communication (SemCom) must move beyond transmitting compact representations toward coordinating reusable skill states under channel constraints.} Existing SemCom aims to transmit task-relevant information rather than reconstructing every source symbol~\cite{xie2021deep,zhang2024unified,qin2024ai}. Deep joint source-channel coding (DeepJSCC) maps visual sources directly into channel symbols and achieves graceful degradation over noisy channels~\cite{bourtsoulatze2019deep}, while recent task-adaptive SemCom introduces prompts, conditional rate-distortion objectives, or tokenized representations to guide semantic coding~\cite{qiao2025token,he2026taskadaptive, zhang2026promptguided}. {Although these methods advance semantic feature coding, prompt conditioning, and tokenized transmission, they mainly optimize each transmission instance independently and do not explicitly exploit reusable execution states across ordered visual workflows.}

A high-level visual intent may unfold as an ordered workflow or modular visual program~\cite{gupta2023visual}, e.g., detecting a person, segmenting the detected instance, and estimating its keypoints~\cite{he2017mask}. In such workflows, later stages inherit execution states such as bounding boxes, masks, or intermediate token memory, and therefore require mainly state-relevant semantic updates rather than another full-token transmission~\cite{li2026SemanticVLA}. Repeatedly sending the full token grid wastes wireless resources on redundant background and already available context, while compressing a compound instruction into a single prompt may cause task interference and attention drift. Therefore, {the key gap is a workflow-level SemCom mechanism that uses skill states to decide which tokens should be transmitted, which context can be reused, and how task-ready representations should be reassembled at the receiver.}


To address this challenge, this letter proposes \textbf{SkillComm}, a skill-driven SemCom framework for workflow-aware token prioritization. To the best of our knowledge, this is the first work that introduces skill-driven workflow control into visual SemCom. As summarized in Tab.~\ref{tab:comparison}, SkillComm organically couples skill-based workflow control with workflow-aware token prioritization, memory-assisted reassembly, and visual JSCC. SkillComm decomposes a high-level visual intent into an executable skill workflow synchronized by a shared \textit{Skill-Book}. Unlike prompt-conditioned SemCom that mainly treats task descriptions as conditioning signals, SkillComm uses reusable skill states as communication context to prioritize state-active tokens before JSCC transmission and to restore task-ready token grids through memory-assisted reassembly. 

{By advancing visual SemCom from prompt-conditioned full-token delivery to workflow-aware semantic information delivery, SkillComm makes the following contributions.}

\begin{itemize}
    \item \textbf{{Skill-Driven SemCom Architecture.}} 
    {We propose {SkillComm}, a visual SemCom framework for sequential workflows. A shared \textit{Skill-Compiler} and \textit{Skill-Book} decompose intents into executable skills, enabling coordinated workflow execution across the wireless link.}

    \item \textbf{{Workflow-Aware Token Prioritization.}} 
    {A memory-driven selection mechanism exploits cross-step states to transmit only active semantic innovations. The receiver then reconstructs task-ready token grids by merging decoded features with cached context.}


    \item \textbf{{Experimental Validation.}} {On the MS COCO 2017 workflow, SkillComm achieves a \textbf{51.2\%} token transmission cost reduction and retains \textbf{99.4\%} upper-bound-normalized AP at 15 dB Rayleigh fading.}
\end{itemize}

\begin{table}[!t]
\centering
\caption{Comparison with representative approaches.}
\label{tab:comparison}
\renewcommand{\arraystretch}{1.12}
\setlength{\tabcolsep}{1.8pt}
\scriptsize
\resizebox{\columnwidth}{!}{
\begin{tabular}{l c c c c c}
\hline
\textbf{Method} 
& \textbf{Year}
& \makecell[c]{\textbf{Workflow}\\\textbf{Compiler}} 
& \makecell[c]{\textbf{Multi-Task}\\\textbf{Control}} 
& \makecell[c]{\textbf{Token }\\\textbf{Prioritization}} 
& \makecell[c]{\textbf{Visual}\\\textbf{JSCC}} \\
\hline

DeepJSCC~\cite{bourtsoulatze2019deep} 
& 2019 & $\times$ & \textemdash & $\times$ & \checkmark \\

TokenCom~\cite{qiao2025token} 
& 2025 & $\times$ & \textemdash & $\times$ & $\times$ \\


SkillCom\textsuperscript{a}~\cite{fu2026skillcom} 
& 2026 & $\triangle$ &  \textemdash  & $\times$ & $\times$ \\

TASC-f~\cite{he2026taskadaptive} 
& 2026 & $\times$ & Prompt & $\times$ & \checkmark \\

PGMT-SC~\cite{zhang2026promptguided} 
& 2026 & $\triangle$ & Prompt & $\times$ & \checkmark \\

\textbf{SkillComm} 
& \textbf{Ours} & \checkmark & Skill & \checkmark & \checkmark \\
\hline
\end{tabular}
}
\\[0.5mm]
\parbox{\columnwidth}{\raggedright
\scriptsize{
$\triangle$ denotes partial support. 
\textsuperscript{a}SkillCom studies text-only LLM-based SemCom, whereas SkillComm targets visual workflow-level SemCom with token prioritization.
}
}
\end{table}

\section{System Model and Problem Formulation}
\label{sec:system_model}

\begin{figure}[!t]
    \centering
    \includegraphics[width=\columnwidth]{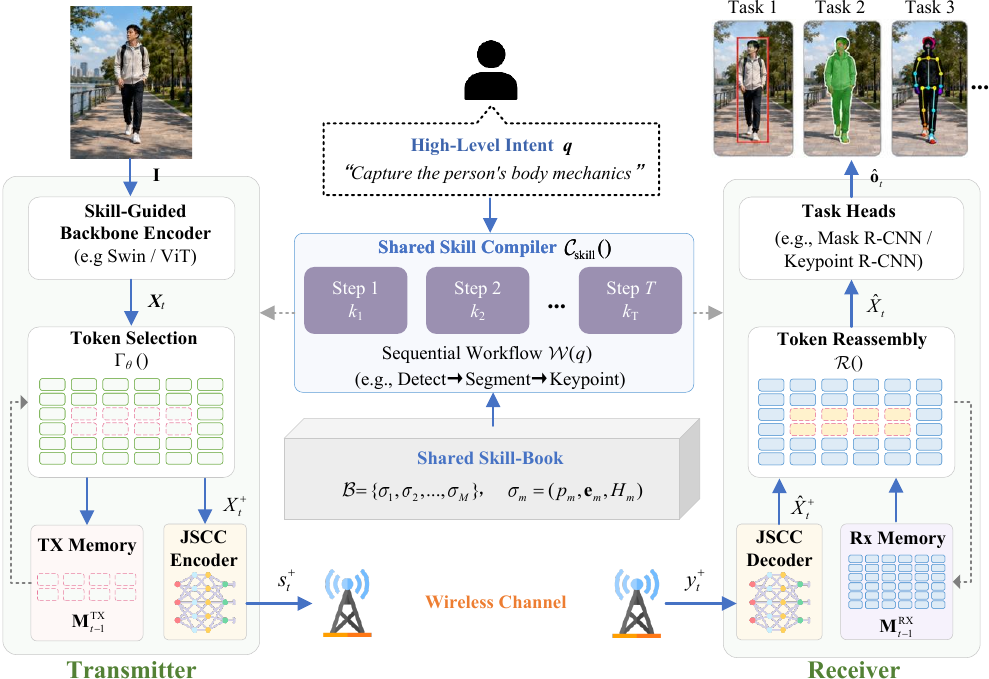}
    \caption{System model of SkillComm. A high-level instruction is decomposed into a synchronized sequential workflow via skill compiler. Reusable skill states guide workflow-aware token prioritization for JSCC transmission, while memory-assisted reassembly reconstructs task-ready token grids at the receiver.}
    \label{fig:skillcomm_system_model}
\end{figure}

Fig.~\ref{fig:skillcomm_system_model} presents the proposed SkillComm architecture. We consider a wireless visual semantic communication system where a transmitter observes an image $\mathbf{I} \in \mathbb{R}^{H \times W \times 3}$ and has a high-level instruction $q\in\mathcal{Q}$, e.g., ``Capture the person's body mechanics.'' The receiver is expected to execute an ordered visual skill workflow, such as \textit{Detect}$\rightarrow$\textit{Segment}$\rightarrow$\textit{Keypoint}, under wireless channel constraints. 

\subsection{Workflow Decomposition using Skill Compiler}
\label{subsec:skill_compiler}

SkillComm assumes that the transmitter and receiver share a synchronized compact \textit{Skill-Book},
\begin{equation}
    \mathcal{B}=\{\sigma_1,\sigma_2,\ldots,\sigma_M\},
\label{eq:skill_book}
\end{equation}
where each reusable visual skill is represented as
\begin{equation}
    \sigma_m=(p_m,\mathbf{e}_m,H_m).
\label{eq:skill_tuple}
\end{equation}
Here, $p_m$ is the textual skill description, $\mathbf{e}_m\in\mathbb{R}^{d_e}$ is the compact skill embedding, and $H_m(\cdot)$ denotes the associated task head, e.g., object detection, instance segmentation, or keypoint estimation. The Skill-Book can be pre-loaded from a library or optimized offline before transmission. In this work, $\mathbf{e}_m$ is instantiated as a single feature vector, while it can be extended to multiple embedding tokens for more complex composite skills.

Given a high-level instruction $q$, the skill compiler maps it to an ordered workflow:
\begin{equation}
    \mathcal{W}(q)=\mathcal{C}_{\mathrm{skill}}(q,\mathcal{B})
    =\left[k_1,k_2,\ldots,k_T\right],
\label{eq:skill_workflow}
\end{equation}
where $k_t\in\{1,\ldots,M\}$ denotes the skill index executed at stage $t$. We define $\mathbf{e}_t \triangleq \mathbf{e}_{k_t}$ and $H_t \triangleq H_{k_t}$ as the scheduled skill embedding and task head. Since both $\mathcal{B}$ and $\mathcal{W}(q)$ are synchronized, the receiver activates the same skill at each stage without repeatedly receiving natural-language prompts.

\subsection{Token Selection Criteria and JSCC Encoding}
\label{subsec:system_token_transmission}

At the $t$-th skill stage, the transmitter extracts a skill-conditioned token canvas:
\begin{equation}
    \mathbf{X}_t = E_{\phi}(\mathbf{I},\mathbf{e}_t)
    =[\mathbf{x}_{t,1},\ldots,\mathbf{x}_{t,N}] \in\mathbb{R}^{N\times d},
\label{eq:system_tokens}
\end{equation}
where $N$ and $d$ denote the spatial token count and token dimension. To avoid repeated full-token transmission, SkillComm selects a state-active token subset $\mathcal{I}_t^{+}\subseteq\{1,\ldots,N\}$ according to
\begin{equation}
    \mathcal{I}_t^{+}=\Gamma_\theta(\mathbf{X}_t,\mathbf{M}_{t-1}^{\mathrm{Tx}},\mathbf{e}_t,\gamma_t),
\label{eq:system_selection}
\end{equation}
where $\mathbf{M}_{t-1}^{\mathrm{Tx}}\in\mathbb{R}^{N\times d}$ is the transmitter-side execution memory and $\gamma_t$ is the normalized channel SNR. For the initial stage, $\mathcal{I}_1^{+}=\{1,\ldots,N\}$ is used to initialize the workflow memory.

The selected tokens are gathered as
$\mathbf{X}_t^{+}=\mathrm{Gather}(\mathbf{X}_t,\mathcal{I}_t^{+})$
and mapped to channel symbols by a JSCC encoder:
\begin{equation}
    \mathbf{s}_t^{+}=\Phi_{\mathrm{enc}}(\mathbf{X}_t^{+}).
\label{eq:system_jscc_enc}
\end{equation}
The skill embedding $\mathbf{e}_t$ conditions the backbone feature extraction, while JSCC operates only on the selected active tokens. The spatial indices are encoded as $\mathbf{b}_t^{\mathrm{idx}}=\mathrm{IndexEnc}(\mathcal{I}_t^{+})$ and transmitted through a reliable control channel. The wireless channel is modeled as
\begin{equation}
    \mathbf{y}_t^{+}=h_t\mathbf{s}_t^{+}+\mathbf{n}_t,
\label{eq:system_channel}
\end{equation}
where $h_t$ is the complex channel coefficient and $\mathbf{n}_t\sim\mathcal{CN}(0,\gamma_{t}^{2}\mathbf{I})$ represents additive white Gaussian noise. The transmitter then updates its local memory as $\mathbf{M}_t^{\mathrm{Tx}}=\mathbf{X}_t$.

\subsection{JSCC Decoding and Token Reassembly Objective}
\label{subsec:receiver_execution}

Upon receiving the wireless signals, the receiver decodes the spatial indices $\hat{\mathcal{I}}_t^{+}=\mathrm{IndexDec}(\mathbf{b}_t^{\mathrm{idx}})$ and applies the JSCC decoder to recover the high-priority active tokens:
\begin{equation}
    \hat{\mathbf{X}}_t^{+}=\Phi_{\mathrm{dec}}(\mathbf{y}_t^{+}).
\label{eq:system_jscc_dec}
\end{equation}
A full token canvas $\hat{\mathbf{X}}_t \in\mathbb{R}^{N\times d}$ is then reconstructed by a reassembly operator $\mathcal{R}(\cdot)$, which scatters the decoded active tokens back to their original grid coordinates and replenishes the unselected slots utilizing the synchronized historical receiver memory $\mathbf{M}_{t-1}^{\mathrm{Rx}}$:
\begin{equation}
    \hat{\mathbf{X}}_t=\mathcal{R}(\hat{\mathbf{X}}_t^{+},\hat{\mathcal{I}}_t^{+},\mathbf{M}_{t-1}^{\mathrm{Rx}}).
\label{eq:system_reassembly}
\end{equation}
Due to the stochastic nature of the noisy wireless channel, $\mathbf{M}_{t-1}^{\mathrm{Tx}} \neq \mathbf{M}_{t-1}^{\mathrm{Rx}}$ generally holds true. Therefore, SkillComm operates robustly without requiring strictly identical memory states or receiver feedback. Because $\hat{\mathbf{X}}_t$ strictly preserves the original grid shape, the activated task head directly executes inference:
\begin{equation}
    \hat{\mathbf{o}}_t=H_t(\hat{\mathbf{X}}_t).
\label{eq:system_output}
\end{equation}
Finally, the receiver updates its memory as $\mathbf{M}_t^{\mathrm{Rx}}=\hat{\mathbf{X}}_t$.


Let $\mathbf{o}_t$ be the ground-truth target for the $t$-th skill and $d_t(\cdot,\cdot)$ be the task distortion metric. SkillComm aims to minimize the cumulative workflow distortion subject to a total communication cost constraint:
\begin{equation}
\resizebox{0.7\columnwidth}{!}{$
    \min_{\Theta} \mathbb{E}\left[ \sum_{t=1}^{T}\alpha_t d_t(\mathbf{o}_t,\hat{\mathbf{o}}_t) +\lambda R_{\mathrm{total}} \right],
$}
\label{eq:system_objective}
\end{equation}
where $\Theta$ encompasses the parameters of the shared feature backbone, JSCC codecs, task heads, and gating modules; $\alpha_t$ balances task priorities, and $\lambda$ dictates the accuracy-rate tradeoff. Here, $R_{\mathrm{total}}$ represents the cumulative transmission rate across the entire sequence of skills. 



\section{Skill-Driven Semantic Communication}
\label{sec:token_selection}

\begin{figure}[!t]
    \centering
    \includegraphics[width=\columnwidth]{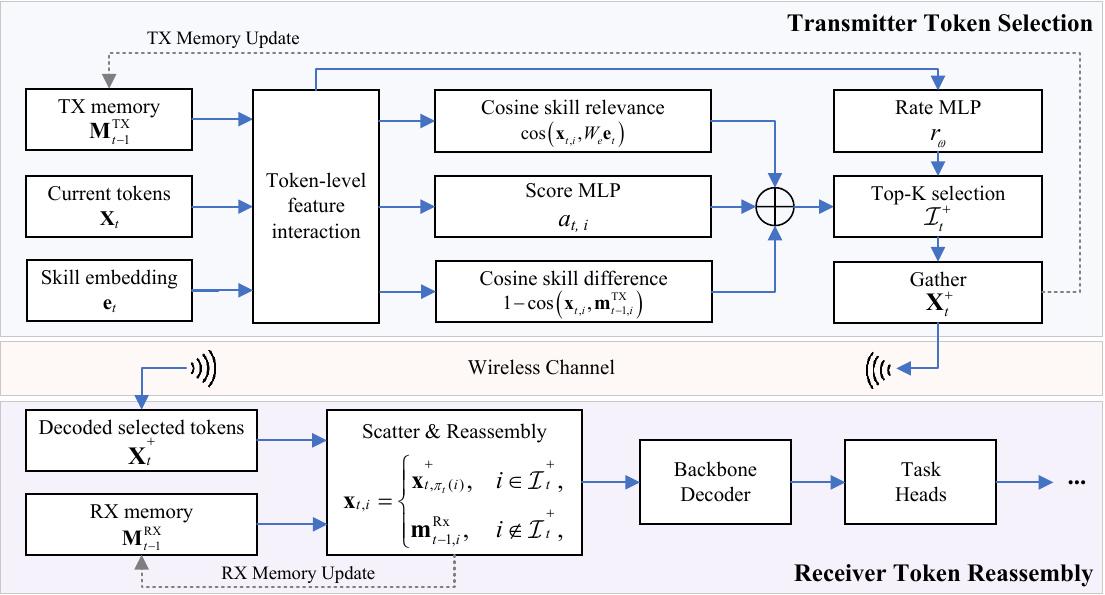}
    \caption{Skill-driven token prioritization and memory-assisted reassembly in SkillComm.}
    \label{fig:skillcomm_token_admission}
\end{figure}

This section realizes the token selection policy $\Gamma_\theta(\cdot)$ and the reassembly operator $\mathcal{R}(\cdot)$ introduced in Sec.~\ref{sec:system_model}. 
{As shown in Fig.~\ref{fig:skillcomm_token_admission}, SkillComm implements workflow-aware token prioritization through memory-aware token scoring, adaptive Top-$K$ JSCC transmission, and memory-assisted token-grid reassembly, enabling state-active semantic innovations to be delivered across sequential skill stages.}



\subsection{Token Scoring at Transmitter}
\label{subsec:transformer_token_sender}
To facilitate cross-task information reuse without redundant transmission, both the transmitter and receiver cache their respective historical states. The transmitter memory is defined as $\mathbf{M}_{t-1}^{\mathrm{Tx}}=[\mathbf{m}_{t-1,1}^{\mathrm{Tx}},\ldots,\mathbf{m}_{t-1,N}^{\mathrm{Tx}}]$, where each memory token $\mathbf{m}_{t-1,i}^{\mathrm{Tx}}$ is initialized as the corresponding feature $\mathbf{x}_{t-1,i}$ generated at the previous stage $t-1$. Similarly, the receiver maintains a local memory $\mathbf{M}_{t-1}^{\mathrm{Rx}}$, where each token $\mathbf{m}_{t-1,i}^{\mathrm{Rx}}$ stores the previously reconstructed feature $\hat{\mathbf{x}}_{t-1,i}$.

A cosine-similarity heuristic can determine whether a current token resembles the historical memory, but fails to differentiate between task-irrelevant background and task-critical novel features. To resolve this, SkillComm introduces a delta-aware token gate. For the $i$-th token, a lightweight Multi-Layer Perceptron (MLP) first evaluates a routing logit:
\begin{equation}
\resizebox{0.88\columnwidth}{!}{$
    a_{t,i}=g_{\theta}\!\left(
    \mathbf{x}_{t,i},
    \mathbf{m}_{t-1,i}^{\mathrm{Tx}},
    \mathbf{e}_t,
    |\mathbf{x}_{t,i}-\mathbf{m}_{t-1,i}^{\mathrm{Tx}}|,
    \mathbf{x}_{t,i}\odot\mathbf{m}_{t-1,i}^{\mathrm{Tx}}
    \right)
$}
\label{eq:token_gate_mlp}
\end{equation}
where $\odot$ denotes element-wise multiplication. The definitive routing score synergizes this logit with explicit current-skill relevance and temporal difference priors:
\begin{equation}
\resizebox{0.88\columnwidth}{!}{$
    s_{t,i}=a_{t,i} +\alpha\cos(\mathbf{x}_{t,i},W_e\mathbf{e}_t) +\beta\left[1-\cos(\mathbf{x}_{t,i},\mathbf{m}_{t-1,i}^{\mathrm{Tx}})\right],
$}
\label{eq:hybrid_gate_score}
\end{equation}
where $W_e$ is a learnable linear projection that maps the skill embedding to the token feature dimension. A rate MLP $r_\omega$ receives average-pooled current, memory, and absolute-difference features, together with $\mathbf{e}_t$ and normalized SNR $\gamma_t$, and predicts
\begin{equation}
\resizebox{0.88\columnwidth}{!}{$
\rho_t=\rho_{\min}+(\rho_{\max}-\rho_{\min})
 \operatorname{sigmoid}\!\left(r_\omega(\mathbf{X}_t,\mathbf{M}_{t-1}^{\mathrm{Tx}},\mathbf{e}_t,\gamma_t)\right),
$}
\label{eq:adaptive_rate_controller}
\end{equation}
where $\rho_t$ is the retained-token fraction and $(\rho_{\min},\rho_{\max})=(0.2,0.8)$. The transmitted token count and indices are
\begin{equation}
 K_t=\operatorname{round}(\rho_tN),\qquad
 \mathcal{I}_t^{+}=\mathrm{TopK}(\{s_{t,i}\}_{i=1}^{N},K_t).
\label{eq:memory_topk}
\end{equation}
Because hard Top-$K$ is non-differentiable, training uses the straight-through mask $\widetilde{\mathbf m}_t=\operatorname{STTopK}(\mathbf s_t,K_t)\in[0,1]^N$. Its forward value is the binary Top-$K$ mask, while its backward derivative is provided by a soft sigmoid relaxation. Training then minimizes
\begin{equation}
\resizebox{0.55\columnwidth}{!}{$
 \mathcal{L}=\sum_{t=1}^{T}\mathcal{L}_t
 +\lambda\sum_{t=2}^{T}\frac{\|\widetilde{\mathbf m}_t\|_1}{N},
$}
\label{eq:route_loss}
\end{equation}
where $\mathcal{L}_t$ is the downstream loss for skill $t$ and $\lambda$ controls the accuracy--rate tradeoff. Thus, inference still transmits exactly $K_t$ tokens; $\widetilde{\mathbf m}_t$ exists only to train the hard Top-$K$ token selector.

\subsection{Memory-Assisted Token Reassembly at Receiver}
\label{subsec:gather_transmit_merge}

Following token gathering and JSCC transmission of the active subset $\mathbf{X}_t^{+}$, the receiver performs cross-step feature reassembly. Because the receiver lacks access to the un-transmitted portions of the transmitter's current state, it cannot execute dynamic fusion mechanisms based on $\mathbf{X}_t$. Instead, the receiver gracefully overwrites the selected positions with the newly arrived tokens and directly inherits the unselected positions from its local historical memory:
\begin{equation}
    \hat{\mathbf{x}}_{t,i}=\begin{cases} \hat{\mathbf{x}}_{t,\pi_t(i)}^{+}, & i\in\hat{\mathcal{I}}_t^{+},\\ \mathbf{m}_{t-1,i}^{\mathrm{Rx}}, & i\notin\hat{\mathcal{I}}_t^{+}, \end{cases}
\label{eq:token_merge}
\end{equation}
where $\pi_t(i)$ is an order-preserving mapping from the original global index $i$ to its compact position in the transmitted sequence $\mathbf{X}_t^{+}$. This pure memory-overwrite mechanism ensures robust feature restoration without introducing supplementary communication overhead.


For a workflow spanning $T$ stages, a prompt-guided full-token benchmark requires the transmission of $TN$ tokens. In contrast, SkillComm incurs full communication cost only at the initial stage, yielding a drastically reduced cumulative workflow token utilization of:
\begin{equation}
    R_{\mathrm{total}} =N+\sum_{t=2}^{T} K_t.
\label{eq:harness_cost_detail}
\end{equation}
Concurrently, the index transmission overhead is bounded by $B_{\mathrm{idx},t} = \lceil \log_2 \binom{N}{K_t} \rceil$. This analytical formulation explicitly guarantees that by dynamically modulating $K_t$ via the rate controller, SkillComm effectively eliminates the repeated transmission of redundant semantic tokens.



\section{Experimental Results}
\label{sec:simulation}

This section evaluates SkillComm on a representative compound intent, ``Capture the person's body mechanics,'' instantiated as $\mathrm{Detect}\!\rightarrow\!\mathrm{Segment}\!\rightarrow\!\mathrm{Keypoint}$: localizing the person, delineating the body, and estimating the joints. {This workflow is used as a representative benchmark to validate sequential skill execution, while the proposed mechanism is not tied to this specific task chain.} We report stage-specific intent conditioning, workflow-level token efficiency, and atomic-task AP$_{\rm box}$, AP$_{\rm mask}$, and AP$_{\rm kp}$.

\subsection{Experimental Setup}
\label{subsec:setup}

Experiments are conducted on 100 person-containing images randomly sampled from the MS COCO 2017 validation set~\cite{lin2014microsoft}. A prompt-conditioned Swin Transformer processes $448\times448$ images and produces a $14\times14\times256$ token canvas, yielding $N=196$ spatial tokens per stage. Pre-trained Mask R-CNN and Keypoint R-CNN heads consume the reassembled canvas for detection, segmentation, and keypoint estimation. Therefore, a conventional full-token execution of the $\mathrm{Detect} \rightarrow \mathrm{Segment} \rightarrow \mathrm{Keypoint}$ workflow requires transmitting $3N=588$ tokens.

During training, the delta-aware ranking gate is optimized for 10,000 iterations. The SNR-aware rate controller is trained for 5,000 iterations with the backbone and task heads frozen, followed by 500 Rayleigh-calibration iterations. The training SNR is continuously sampled from $[0,15]$ dB. For evaluation, Rayleigh fading is applied at $\{0,3,6,9,12,15\}$ dB with shared channel realizations across methods. 

We compare SkillComm with the following references and ablation variants:
\begin{itemize}
    \setlength{\itemsep}{0pt}
    \setlength{\parskip}{0pt}
    \setlength{\parsep}{0pt}

    \item \textbf{NC-UB:} The no-channel upper bound, where the full workflow is executed locally without wireless degradation.

    \item \textbf{TASC-f~\cite{he2026taskadaptive}:} A task-adaptive SemCom reference based on prompt/conditional transmission, included only for applicable comparisons since it does not provide automatic workflow compilation or workflow-level token reuse.

    \item \textbf{PGMT-SC~\cite{zhang2026promptguided}:} A single-task prompt-guided SemCom reference. For a compound intent, it executes one favorable atomic task and is therefore excluded from per-task AP curves.

    \item \textbf{SkillComm-Full:} The sequential baseline, where task prompts are manually supplied step by step and all $N=196$ tokens are transmitted at each stage.
    
    \item \textbf{SkillComm-Fixed:} An ablation that retains automatic workflow compilation and execution memory, but uses deterministic cosine Top-$K$ selection with a fixed $50\%$ downstream token ratio.
    
    \item \textbf{SkillComm-Adaptive:} The complete method with automatic skill compilation, cross-step memory, learned delta-aware token gating, and adaptive SNR-aware rate control.
\end{itemize}

\subsection{Stage-Conditioning Alignment and Token Utilization}
\label{subsec:workflow_evaluation}

This sub-section evaluates the overall intention compiler performance as well as token savings. Since TASC-f requires explicit task feedback, we instantiate its no-feedback adaptation by selecting one known atomic prompt through frozen multilingual MiniLM similarity~\cite{wang2020minilm}. The intent has similarities $[0.447,0.319,0.357]$ to detection, segmentation, and keypoint prompts, respectively, and therefore selects detection for the entire request. Following the mixture-of-prompts (MOP) protocol of PGMT-SC, its unseen-intent embedding is $0.30\mathbf e_{\rm det}+0.30\mathbf e_{\rm seg}+0.40\mathbf e_{\rm kp}$. SkillComm first compiles the intent and then switches to the corresponding atomic embedding at each workflow stage.

For the compound intent, let $\mathbf{w}_t$ denote the effective routing weights over the atomic prompts at stage $t$, and let $\mathbf{w}_t^\star$ denote the target one-hot routing vector of the required skill. We measure \textit{stage-conditioning alignment} (SCA) using a normalized Brier-style score~\cite{guo2017calibration}:
\begin{equation}
\resizebox{0.60\columnwidth}{!}{$
    \mathrm{SCA}=1-\frac{1}{2T}\sum_{t=1}^{T}
    \left\|\mathbf{w}_t-\mathbf{w}_t^\star\right\|_2^2.
$}
\label{eq:sca}
\end{equation}
For the target D$\rightarrow$S$\rightarrow$K workflow, the desired routing vectors are $\mathbf{w}_1^\star=(1,0,0)$, $\mathbf{w}_2^\star=(0,1,0)$, and $\mathbf{w}_3^\star=(0,0,1)$. Both $\mathbf{w}_t$ and $\mathbf{w}_t^\star$ lie on the probability simplex, whose maximum squared Euclidean distance is two, e.g., $\|(1,0,0)-(0,1,0)\|_2^2=2$. Hence, the factor $2T$ normalizes SCA to $[0,1]$: a score of one means that every stage receives its correct atomic condition, while a mismatch decreases the score according to its full routing-vector distance. This score is evaluated in the discrete skill-routing space rather than by cosine similarity between prompt embeddings. A hard-selected prompt has one-hot routing weights, whereas PGMT-SC uses its normalized MOP coefficients directly.

\begin{table}[t]
\centering
\caption{Stage-conditioning alignment and workflow-level token efficiency.}
\label{tab:workflow_compilation_tokens}
\renewcommand{\arraystretch}{1.08}
\setlength{\tabcolsep}{1.4pt}
\scriptsize
\resizebox{\columnwidth}{!}{%
\begin{tabular}{lcccc}
\hline
Method & Effective Skill Sequence  & SCA & Tokens$^\dagger$ & Saving \\
\hline
NC-UB & D$\rightarrow$S$\rightarrow$K (channel-free reference) & 1 & -- & -- \\
TASC-f$^\dagger$~\cite{he2026taskadaptive} & D$\rightarrow$D$\rightarrow$D & 33.3\% & 588.0 & 0.0\% \\
PGMT-SC$^\dagger$~\cite{zhang2026promptguided} & M$\rightarrow$M$\rightarrow$M & 66.3\% & 588.0 & 0.0\% \\
SkillComm-Full & D$\rightarrow$S$\rightarrow$K & 100.0\% & 588.0 & 0.0\% \\
SkillComm-Fixed & D$\rightarrow$S$\rightarrow$K & 100.0\% & 392.0 & 33.3\% \\
\textbf{SkillComm-Adaptive} & \textbf{D$\rightarrow$S$\rightarrow$K} & \textbf{100.0\%} & \textbf{287.1} & \textbf{51.2\%} \\
\hline
\end{tabular}}
\\[0.2mm]
\parbox{\columnwidth}{\scriptsize D, S, and K denote detection, segmentation, and keypoint prompts; M denotes the fixed MOP weights $(0.30,0.30,0.40)$. \\
$^\dagger$TASC-f and PGMT-SC are evaluation-unrolled over three task heads only to assess stage-conditioning mismatch under the same workflow, without native workflow execution.
}
\end{table}

Tab.~\ref{tab:workflow_compilation_tokens} separates stage-wise intent resolution from transmission cost for the same compound intent. TASC-f repeats $\mathbf{w}_t=(1,0,0)$ at all stages; its three squared errors are $0$, $2$, and $2$, giving $1-4/6=33.3\%$ SCA. PGMT-SC repeats $\mathbf{w}_t=(0.30,0.30,0.40)$; its stage-wise errors are $0.74$, $0.74$, and $0.54$, giving $1-2.02/6=66.3\%$. This higher value reflects partial soft alignment, but not stage-specific decomposition. SkillComm compiles the intent into the correct one-hot sequence, incurs zero routing error, and obtains 100\% SCA. Based on this workflow, its learned gate uses 287.1 rather than 588 tokens, leading to a token saving of 51.2\%.

\begin{figure*}[t]
    \centering
    \subfloat[Stage 1: Object Detection]{\includegraphics[width=0.3\textwidth]{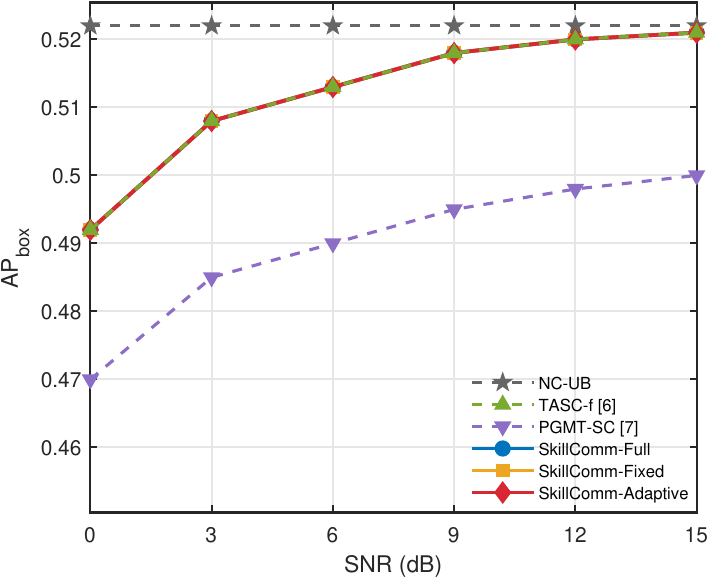}}
    \hfill
    \subfloat[Stage 2: Instance Segmentation]{\includegraphics[width=0.3\textwidth]{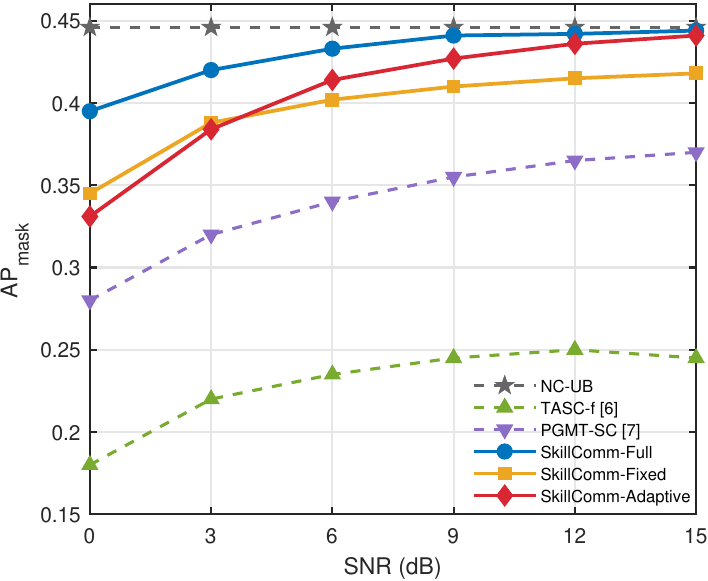}}
    \hfill
    \subfloat[Stage 3: Keypoint Detection]{\includegraphics[width=0.3\textwidth]{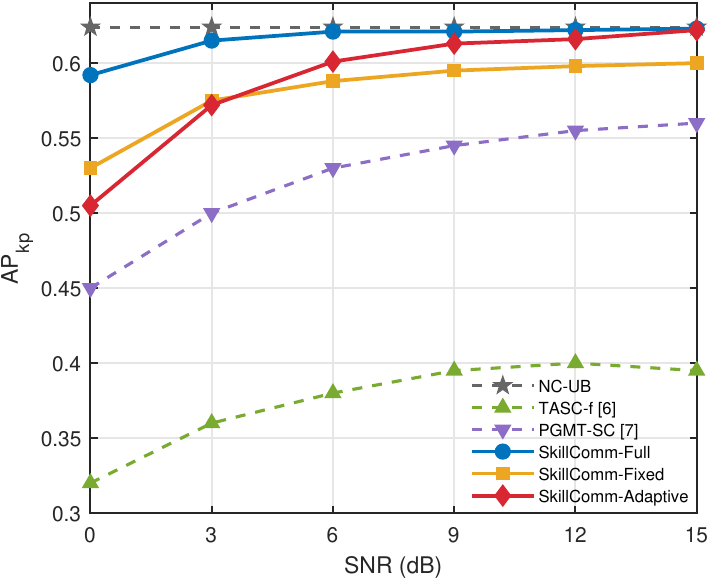}}
    \caption{Per-task COCO Average Precision (AP) over Rayleigh fading channels for the \textit{Detect}$\rightarrow$\textit{Segment}$\rightarrow$\textit{Keypoint} workflow.}
    \label{fig:rayleigh_task_ap}
\end{figure*}

\subsection{Per-Task AP Evaluation over Rayleigh Fading}
\label{subsec:main_comparison}

Fig.~\ref{fig:rayleigh_task_ap} reports the per-task AP of the three workflow stages under Rayleigh fading channels. In the first stage, all SkillComm variants transmit the full token grid and therefore achieve nearly identical detection performance. TASC-f also performs well at this stage because it interprets the compound intent as a detection prompt, whereas PGMT-SC suffers from mixed multi-task conditioning. In the segmentation and keypoint stages, however, TASC-f continues to use the detection condition and degrades significantly, while SkillComm maintains stage-specific execution through the skill compiler.

Among the SkillComm variants, {SkillComm-Full} provides the upper performance reference by transmitting all tokens, whereas {SkillComm-Fixed} and {SkillComm-Adaptive} reduce token usage through selective transmission. SkillComm-Fixed is more robust at low SNR due to its conservative token ratio, while SkillComm-Adaptive surpasses it at high SNR by selecting higher-value tokens according to channel and memory-aware token priorities. Overall, SkillComm-Adaptive achieves 94.3\% average AP retention relative to NC-UB across 0--15 dB and all three tasks, and reaches 99.4\% retention at 15 dB, demonstrating that workflow-aware token prioritization preserves task-critical semantics with substantially lower communication cost.

\section{Conclusions}
\label{sec:conclusion}

This letter proposed {SkillComm}, {a skill-driven SemCom framework that uses reusable skill states to guide workflow-aware token prioritization and memory-assisted token-grid reassembly}. Instead of treating each task as an isolated full-token transmission problem, SkillComm synchronizes ordered skill execution through a shared \textit{Skill-Book}, exploits execution memory to prioritize state-active tokens for JSCC transmission, and restores task-ready token grids at the receiver. {Experiments on the MS COCO 2017 \textit{Detect}$\rightarrow$\textit{Segment}$\rightarrow$\textit{Keypoint} workflow show that SkillComm reduces token transmission cost by 51.2\%, while retaining 99.4\% upper-bound-normalized AP in the high-SNR 15-dB Rayleigh fading regime.} These results validate a general principle for workflow-level SemCom: reusable skill states can guide wireless systems to transmit semantic innovations rather than repeated full representations. {Future work will develop feedback-loop skills that adapt the next skill, token budget, and refinement transmission using receiver-side confidence, channel state information, and memory-reassembly errors.}

\bibliographystyle{IEEEtran}
\bibliography{IEEEabrv, Ref_SkillComm}

\vspace{12pt}
\end{document}